\RequirePackage{ifpdf}
\ifpdf 
\documentclass[pdftex]{sigma}
\else
\documentclass{sigma}
\fi

\begin{document}
\allowdisplaybreaks

\renewcommand{\PaperNumber}{027}

\FirstPageHeading

\ShortArticleName{Integrable Anisotropic Evolution Equations on a
Sphere}

\ArticleName{Integrable Anisotropic Evolution Equations\\ on a
Sphere} 

\Author{Anatoly G. MESHKOV~$^*$ and Maxim Ju. BALAKHNEV}

\AuthorNameForHeading{A.G. Meshkov and M.Ju. Balakhnev}

\Address{Orel State University, 95 Komsomol'skaya Str.,  Orel,
302026 Russia}

\Email{\href{mailto:meshkov@orel.ru}{meshkov@orel.ru},
\href{mailto:maxibal@yandex.ru}{maxibal@yandex.ru}}

\URLaddressMarked{\url{http://www.orel.ru/meshkov}}

\ArticleDates{Received September 25, 2005, in final form December
09, 2005; Published online December 14, 2005}

\Abstract{V.V. Sokolov's modifying symmetry approach  is applied to
anisotropic evolution equations of the third order on the $n$-dimensional sphere.
The main result is a complete classification
of such equations. Auto-B\"acklund transformations are also found for all equations.}


\Classification{35Q58; 35L65; 37K10; 37K35}

\section{Introduction}
 In this paper we are dealing with the problem of the
 symmetry classification of the integrable vector evolution equations
of the third order. Completely integrable equations possess many
remarkable properties and are often interesting for applications.
 As examples of such equations we may point out  two well
 known modified Korteweg--de Vries equations
\begin{gather*}
\boldsymbol u_t = \boldsymbol u_{xxx} + (\boldsymbol u ,\boldsymbol u)\boldsymbol u_x,\qquad
\boldsymbol u_t = \boldsymbol u_{xxx}
+ (\boldsymbol u,\boldsymbol u)\boldsymbol u_x + (\boldsymbol u, \boldsymbol u_x)\boldsymbol u,
\end{gather*}
where $(\cdot\ , \cdot )$ is a scalar product, $\boldsymbol u=(u^1,\dots,u^n)$.
These equations are integrable by the inverse scattering
method for any vector dimension.  Another example of integrable anisotropic evolution equation on the sphere
is the higher order Landau--Lifshitz  equation
\begin{gather}
\boldsymbol u_t=\left(\boldsymbol u_{xx}+{3\over 2} (\boldsymbol u_x,  \boldsymbol u_x)
\boldsymbol u \right)_x +{3\over 2} (\boldsymbol u,
R\boldsymbol u) \boldsymbol u_x,  \qquad (\boldsymbol u,  \boldsymbol u)=1,
\label{landlif}
\end{gather}
where $R$ is a diagonal constant matrix. Complete integrability of this equation was proved in~\cite{golsok}.

To investigate integrability of such equations a modification of the symmetry approach was proposed in~\cite{sokw}.
Different examples of integrable vector equations can also be  found in~\cite{sokw}.
Sokolov's method can be applied to any vector equation
\begin{gather}
\boldsymbol u_t=f_n\boldsymbol u_{n}+ f_{n-1} \boldsymbol u_{n-1}+\cdots + f_0 \boldsymbol u \label{nth}
\end{gather}
with scalar coefficients $f_i$.
Henceforth $\boldsymbol u=(u^1,\dots, u^{n+1})$ denotes an unknown vector,
 $\boldsymbol u_{k}=\partial ^k\boldsymbol u/\partial x^k$ and
$f_i$ are some scalar functions  depending on  the scalar products
$(\boldsymbol u_i, \boldsymbol u_j)$, $i\leqslant j$. Moreover, dependence
of $f_i$ on more than one scalar products $(\cdot, \cdot )_i$,
$i=1,2,\dots$ may be introduced. It is clear that any equation (\ref{nth}) with the
Euclidean scalar product is invariant with respect to an arbitrary
constant orthogonal transformation of the vector $\boldsymbol u$.
Therefore the equation (\ref{nth}) with a  unique scalar product
is called isotropic. When $f_i$ depend on two or more scalar products
$(\boldsymbol u_i, \boldsymbol u_j)_k$, we call the equation (\ref{nth})
anisotropic. The scalar products may have a different  nature. Only
two properties of the scalar products are essential for us  --- bilinearity
and continuity. Vector $\boldsymbol u$ may be both real and complex.

The symmetry approach \cite{soksh,MSY,mshsok,fokas,adshyam}
is based on the observation that all integrable evolution
equations with one spatial variable possess local higher
symmetries or, which is the same, higher commuting flows.
Canonical conserved densities $\rho_i$,
$i=0,1,\dots$ make the central notion of this approach. These densities can be expressed in
terms of the coefficients of the equation under consideration.
The evolutionary derivative of  $\rho_i$ must be the total $x$-derivative of
some local function $ \theta_i$:
\begin{gather}
D_t \rho_i(u) = D_x \theta_i(u),\qquad i=0,1,\dots . \label{cond}
\end{gather}
It follows from (\ref{cond}) that the variational derivative
$\delta D_t \rho_i(u)/\delta u$ is zero. Both equations $\delta D_t \rho_i/\delta u$ $=\,0$,
$i=0,1,\dots$ and (\ref{cond}) are called the integrability conditions.

The modified symmetry approach has been recently applied to the equations on $\mathbb S^n$~\cite{M-S1}
and on $\mathbb R^n$ \cite{M-S2,Bal}, where the third order equations in the form
\begin{gather}
\boldsymbol u_t=\boldsymbol u_{xxx}+f_2\boldsymbol u_{xx}+f_1\boldsymbol u_x+f_0\boldsymbol u,    \label{3th}
\end{gather}
were classified under several restrictions: $(\boldsymbol u, \boldsymbol u) =1$ in~\cite{M-S1},
$\boldsymbol u_t=(\boldsymbol u_{xx}+f_1\boldsymbol u_x+f_0\boldsymbol u)_x$ in~\cite{M-S2} and $f_2=0$ in~\cite{Bal}.
We have also tried to  classify the equations (\ref{3th}) on $\mathbb R^n$ but the attempt
has failed because of great
computational difficulties. Moreover, the integrable equations that we found
proved to be too cumbersome to be applied
to any scientific problem. That is why the restrictions were used
in the above-mentioned articles.  In contrast with  $\mathbb R^n$
the list of the anisotropic  integrable equations on  $\mathbb S^n$
presented in Section~2 is short and contains  at the least
one interesting equation~(\ref{EQ1}).

Thus, the subject of the article may be defined
as the symmetry classification of the equations~(\ref{3th}) with the constraint $(\boldsymbol u,
\boldsymbol u)\equiv\boldsymbol u^2 =1$. The coefficients
$f_i$ are assumed to be depending on both isotropic variables $u_{[i,j]}$ and
anisotropic variables $\tilde u_{[i,j]}$
\begin{gather}
u_{[i,j]}=(\boldsymbol u_i, \boldsymbol u_j),\qquad
\tilde u_{[i,j]}=\langle\boldsymbol u_i, \boldsymbol u_j\rangle, \qquad i\leqslant j   \label{vars}
\end{gather}
with $i\leqslant j\leqslant 2$. The constraint $\boldsymbol u^2 =1$ implies
\begin{gather}
(\boldsymbol u, \boldsymbol u_t)=0, \qquad u_{[0,1]}=0,\quad u_{[0,2]}=-u_{[1,1]},\quad
 u_{[0,3]}=-3u_{[1,2]}, \quad \dots.      \label{cst}
\end{gather}
From these constraints it follows that $f_0=f_2u_{[1,1]}+3u_{[1,2]}$ in the equation~(\ref{3th}).

In this paper we shall consider the equations (\ref{3th}) that are integrable for arbitrary dimension~$n$ of the sphere.
In addition, we assume that the coefficients $f_i$ do not depend on $n$. In virtue of the arbitrariness of $n,$ the variables (\ref{vars})
will be regarded as {\bf independent}. The functional independence of $\{u_{[i,j]},\, \tilde u_{[i,j]},\,
 i\leqslant j\}$ is a crucial
requirement in all our considerations.

It is easy to see that the stereographic projection maps any equation (\ref{3th}) on $\mathbb S^n$ to some
anisotropic equation on $\mathbb R ^{n-1}$.

In Section 2 we present a complete list of integrable anisotropic
equations of the form (\ref{3th}) on the sphere $\mathbb S ^{n}$.
And a scheme of computations  is presented in Section~3.

In order to prove that all equations from the list are really
integrable, we find, in Section~4, an auto-B\"acklund transformation
involving a ``spectral'' parameter for each of the equations.

\section{Classification results}

In this section we formulate some classification statements
 concerning integrable evolution equations of third order on the
$n$-dimensional sphere. This classification problem is much
simpler than the similar problem on $\mathbb R^n$. Indeed,
the set of the independent variables (\ref{vars}) on $\mathbb S^n$
is reduced because of the constraints (\ref{cst}).  It is easy to see
that we can express all variables of the form $u_{[0,k]}$,
$k\geqslant 1$ in terms of the remaining independent scalar products.
So, the complete set of dynamical variables on the sphere is
\begin{gather}\label{dynvar}
\{u_{[i,j]},\  1\leqslant i\leqslant j; \ \tilde u_{[i,j]},\ 0\leqslant i\leqslant j\}.
\end{gather}
Therefore the coefficients of the equation (\ref{3th})
on $\mathbb S^n$ {\it a priori} depend on  nine independent va\-riab\-les
\renewcommand{\theequation}{\arabic{equation}${}'$}
\setcounter{equation}{6}
\begin{gather}
u_{[1,1]},\ u_{[1,2]},\ u_{[2,2]},\ \tilde u_{[0,0]},\ \tilde u_{[0,1]},\ \tilde u_{[0,2]},\
 \tilde u_{[1,1]},\ \tilde u_{[1,2]},\ \tilde u_{[2,2]},
\label{dynvar1}
\end{gather}
whereas in the case of  $\mathbb R^n$ they are functions of twelve variables.

 Let $g_{ij}$ and $\tilde g_{ij}$ be the first and the second metric tensors, $u_{[k,l]}=\sum\limits_{i,j} g_{ij}u^i_k u^j_l,\
\tilde u_{[k,l]}=\sum\limits_{i,j}\tilde g_{ij}u^i_k u^j_l$. Then the equation (\ref{3th}) and the constraint $\boldsymbol u^2=1$ are obviously invariant
under the transformation $\tilde g_{ij}\to \alpha \tilde g_{ij}+\lambda  g_{ij}$ where $\alpha$ and $\lambda$ are constants and $\alpha\ne0$.
It is equivalent to the following transformation of the dependent variables
\renewcommand{\theequation}{\arabic{equation}}
\begin{gather}\label{tr}
\tilde u_{[k,l]}\to \alpha \tilde u_{[k,l]}+\lambda u_{[k,l]}.
\end{gather}
We used this transformation to classify the integrable equations  (\ref{3th}).

\begin{theorem}
 If the anisotropic equation on $\mathbb S^{n}$
\begin{gather}
\boldsymbol u_t=\boldsymbol u_{xxx}
+f_2\boldsymbol u_{xx}+f_1\boldsymbol u_x+(f_2u_{[1,1]}+3u_{[1,2]})\boldsymbol u, \label{EQ}
\end{gather}
 possesses an infinite series of canonical conservation laws
$(\rho_{k})_{t}=(\theta_{k})_{x}$, $k=0,1,2\dots$,
where~$\rho_{k}$ and $\theta_{k}$ are functions of variables \eqref{dynvar},
then this equation can be reduced, with the help of the
transformation \eqref{tr}, to one of the following equations
\begin{gather}\label{EQ1}
\boldsymbol u_t=\boldsymbol u_3+\frac{3}{2}\big(u_{[1,1]}+\tilde u_{[0,0]}\big)
\boldsymbol u_1+3 u_{[1,2]} \boldsymbol u, \\
\label{EQ2}
\boldsymbol u_t=\boldsymbol u_3 - 3\frac{u_{[1,2]}}{u_{[1,1]}}\boldsymbol u_2+\frac{3}{2}\left(\frac{u_{[2,2]}}
{u_{[1,1]}}+\frac{u_{[1,2]}^2}{u_{[1,1]}^2}+
\frac{\tilde u_{[1,1]}}{u_{[1,1]}}\right)\boldsymbol u_1, \\
\label{EQ3}
\boldsymbol u_t=\boldsymbol u_3-3\frac{u_{[1,2]}}{u_{[1,1]}}
\boldsymbol u_2 +\frac{3}{2}\left(\frac{u_{[2,2]}}{u_{[1,1]}}+
\frac{u_{[1,2]}^2}{u_{[1,1]}^2}-\frac{(\tilde u_{[0,1]}
+u_{[1,2]})^2}{(u_{[1,1]}+ \tilde u_{[0,0]}) u_{[1,1]}}
+\frac{\tilde u_{[1,1]}}{u_{[1,1]}}\right)\boldsymbol u_1, \\
\label{EQ4}
\boldsymbol u_t=\boldsymbol u_3-3\frac{\tilde  u_{[0,1]} }{\tilde  u_{[0,0]}} \boldsymbol u_2
-3\left(\frac{2\tilde  u_{[0,2]}+\tilde  u_{[1,1]}+a}{2\tilde  u_{[0,0]}}-\frac{5}{2}
\frac{\tilde  u_{[0,1]}^2 }{\tilde  u_{[0,0]}^2}\right)\boldsymbol u_1\!
+3 \left(u_{[1,2]} -\frac{\tilde  u_{[0,1]} }{\tilde  u_{[0,0]}}u_{[1,1]}\right)\boldsymbol u ,\!\!\!  \\
\label{EQ5}
\boldsymbol u_t=\boldsymbol u_3-3\frac{\tilde  u_{[0,1]} }{\tilde  u_{[0,0]}} \boldsymbol u_2
-3\left(\frac{\tilde  u_{[0,2]} }{\tilde  u_{[0,0]}}-2\frac{ \tilde
u_{[0,1]}^2 }{\tilde  u_{[0,0]}^2}\right)\boldsymbol u_1
+3 \left(u_{[1,2]} -\frac{\tilde  u_{[0,1]} }{\tilde  u_{[0,0]}}u_{[1,1]}\right)\boldsymbol u ,  \\
\boldsymbol u_t= \boldsymbol u_3- 3\frac{\tilde  u_{[0,1]} }{\tilde  u_{[0,0]}}
\left( \boldsymbol u_2+u_{[1,1]}\boldsymbol u \right) +3 u_{[1,2]}\boldsymbol u\nonumber\\
\phantom{\boldsymbol u_t=}{} +\frac{3}{2}\left( -\frac {u_{[2,2]}}{\tilde u_{[0,0]}}+\frac {(u_{[1,2]}+\tilde u_{[0,1]})^{2}}{\tilde u_{[0,0]}
(\tilde u_{[0,0]}+u_{[1,1]})}+\frac {(\tilde u_{[0,0]}+u_{[1,1]})^{2}}{\tilde u_{[0,0]}}
+\frac {\tilde u_{[0,1]}^{2}-\tilde u_{[0,0]} \tilde u_{[1,1]}}{\tilde u_{[0,0]}^{2}}\right) \boldsymbol u_1,
\label{EQ6}\\
\boldsymbol u_t=\boldsymbol u_3+3\left( \frac{\tilde  u_{[0,1]}
\tilde  u_{[0,2]} }{\xi}-\frac{\tilde  u_{[1,2]}\tilde  u_{[0,0]} }{\xi}
+\frac{\tilde  u_{[0,1]} }{\tilde  u_{[0,0]}} \right)\left( \boldsymbol u_2
+u_{[1,1]}\boldsymbol u\right)+3u_{[1,2]}\boldsymbol u \nonumber\\
\phantom{\boldsymbol u_t=}{} +\frac{3}{2\xi^2\tilde  u_{[0,0]}^2 } \Big( \tilde  u_{[0,0]}^3\tilde  u_{[2,2]}
\xi-\xi (\xi+\tilde  u_{[0,2]}\tilde  u_{[0,0]} )^2+(\tilde  u_{[0,0]}^2\tilde  u_{[1,2]}
-2\xi \tilde  u_{[0,1]}\nonumber\\
\phantom{\boldsymbol u_t=}{}
-\tilde  u_{[0,0]}\tilde  u_{[0,1]}\tilde  u_{[0,2]} )^2  \Big)\boldsymbol u_1 -a\frac{\tilde  u_{[0,0]}^2u_{[1,1]}+\tilde  u_{[0,1]}^2}{\tilde  u_{[0,0]}\xi}\boldsymbol u_1,
\qquad \xi=\tilde  u_{[0,0]} \tilde  u_{[1,1]}-\tilde  u_{[0,1]}^2,\label{EQ7}
\\
\boldsymbol u_t=\boldsymbol u_3+3\left( \frac{\tilde  u_{[0,1]}\tilde  u_{[0,2]} }{\xi}
-\frac{\tilde  u_{[1,2]}\tilde  u_{[0,0]} }{\xi}+\frac{\tilde  u_{[0,1]} }{\tilde  u_{[0,0]}} \right)
\left( \boldsymbol u_2+u_{[1,1]}\boldsymbol u\right)+3u_{[1,2]}\boldsymbol u \nonumber\\
\phantom{\boldsymbol u_t=}{} +\frac{3}{\xi } \left( \tilde  u_{[0,0]}\tilde  u_{[2,2]}
-2\tilde  u_{[0,1]} \tilde  u_{[1,2]} -\frac{ (\tilde  u_{[0,2]}\tilde  u_{[0,0]} -2\tilde  u_{[0,1]}^2)(\xi+
\tilde  u_{[0,2]}\tilde  u_{[0,0]})}{\tilde  u_{[0,0]}^2} \right)\boldsymbol u_1, \label{EQ8}\\
\phantom{\boldsymbol u_t=}{} \xi=\tilde  u_{[0,0]} \tilde  u_{[1,1]}-\tilde  u_{[0,1]}^2,\nonumber\\
\boldsymbol u_t=\boldsymbol u_3-3\frac{a\tilde  u_{[0,1]}}{\eta}\boldsymbol u_2
+ 3\frac{u_{[1,2]}\eta-a\tilde  u_{[0,1]}u_{[1,1]} }{\eta}\boldsymbol u+
\frac{3}{2}\left(\frac{\tilde  u_{[2,2]} }{\eta}+\frac{a\xi
- (\tilde  u_{[0,2]}+\eta)^2 }{\eta\tilde  u_{[0,0]}}\right)\boldsymbol u_1\nonumber\\
\phantom{\boldsymbol u_t=}{} +\frac{3}{2}\left(\frac{\big(\tilde  u_{[0,0]}\tilde  u_{[1,2]}-
\tilde  u_{[0,1]}(2a\tilde  u_{[0,0]}+b+\tilde  u_{[0,2]})\big)^2 }{\eta\xi\tilde  u_{[0,0]}}
-b \frac{a \tilde  u_{[0,1]}^2+ \eta\tilde  u_{[0,0]} u_{[1,1]} }
{\eta^2 \tilde  u_{[0,0]}}\right)\boldsymbol u_1,
\label{EQ9}\\
\phantom{\boldsymbol u_t=}{} \eta=a\tilde  u_{[0,0]}+b,
\qquad \xi=\tilde  u_{[0,0]}(\eta-\tilde  u_{[1,1]})+\tilde  u_{[0,1]}^2, \nonumber \\
\boldsymbol u_t=\boldsymbol u_3
+3\left( \frac{\tilde  u_{[0,1]}}{\tilde  u_{[0,0]} }
+\frac{\tilde  u_{[0,1]}\tilde  u_{[0,2]} }{\xi }-\frac{\tilde  u_{[0,0]}\tilde  u_{[1,2]} }{\xi }\right)
 \left(\boldsymbol u_2+u_{[1,1]}\boldsymbol u\right) +3u_{[1,2]}\boldsymbol u \nonumber\\
\phantom{\boldsymbol u_t=}{} +\frac{3}{2}\left(\frac{\tilde  u_{[0,0]}\tilde  u_{[2,2]} }{\xi}+b\tilde
u_{[0,0]}\frac{u_{[1,1]}\eta+a\tilde  u_{[0,1]}^2 }{\eta\xi}
-\frac{(\tilde  u_{[0,0]}\tilde  u_{[0,2]}+\xi)^2 }{\tilde  u_{[0,0]}^2\xi }\right)\boldsymbol u_1\nonumber\\
\phantom{\boldsymbol u_t=}{} +\frac{3}{2}\frac{ \big(  \tilde  u_{[0,0]}^2(a\xi\tilde  u_{[0,1]}-\eta\tilde  u_{[1,2]})
+\eta\tilde  u_{[0,1]}(\tilde  u_{[0,0]}\tilde  u_{[0,2]}+\xi ) \big )^2 }
{\eta \tilde  u_{[0,0]}^2(\xi+\eta)\xi^2 }\boldsymbol u_1,
\label{EQ10}\\
\phantom{\boldsymbol u_t=}{}  \xi=\tilde  u_{[0,0]} \tilde  u_{[1,1]}
-\tilde  u_{[0,1]}^2,\qquad\eta=(a\tilde  u_{[0,0]}+b)\tilde  u_{[0,0]}, \nonumber\\
\boldsymbol u_t= \boldsymbol u_3+\frac{3}{2}\left(\frac{\tilde u_{[0,0]}
\tilde u_{[1,2]}-\tilde u_{[0,1]}\tilde u_{[0,2]}}{\mu(\mu+\tilde u_{[0,0]})}
-2\frac{\tilde u_{[0,1]}}{\mu}\right)(\boldsymbol u_2
+u_{[1,1]}\boldsymbol u)+3u_{[1,2]}\boldsymbol u\nonumber\\
\phantom{\boldsymbol u_t=}{}+\frac{3/2}{\tilde u_{[0,0]}(\mu+\tilde u_{[0,0]})}
\left[\mu^{-2}\big(\tilde u_{[0,0]}\tilde u_{[1,2]}
-\tilde u_{[0,1]}\tilde u_{[0,2]}\big)^2+\tilde u_{[0,0]}
\tilde u_{[2,2]}-\tilde u_{[0,2]}^2\right.\nonumber\\
\phantom{\boldsymbol u_t=}{} - 2\mu^{-2}\tilde u_{[0,1]}\big(\tilde u_{[0,0]}\tilde u_{[1,2]}
-\tilde u_{[0,1]}\tilde u_{[0,2]})(\mu+2\tilde u_{[0,0]})\Big]\boldsymbol u_1
+\big(6\mu^{-2}\tilde u_{[0,1]}^2-3\tilde u_{[0,0]}^{-1}\tilde u_{[0,2]}\big)
\boldsymbol u_1,\nonumber\\
\phantom{\boldsymbol u_t=}{} \mu^2=\tilde u_{[0,1]}^2+\tilde u_{[0,0]}^2-\tilde u_{[0,0]}\tilde u_{[1,1]}. \label{EQ11}
\end{gather}
\end{theorem}

\begin{remark}
The equations (\ref{EQ1})--(\ref{EQ3}) were given in \cite{M-S1};
the equation (\ref{EQ1}) coincides with (\ref{landlif}).
\end{remark}

\begin{remark} Each of the equations (\ref{EQ1})--(\ref{EQ11})
 can contain the term $c\boldsymbol u_1$ in its right-hand side. We removed
these terms by the Galilean transformation as trivial.
The constants $a$ and $b$ are arbitrary. One can set $a=0$ in (\ref{EQ4}) and in (\ref{EQ7}).
In (\ref{EQ9}) we may choose $a=0$ or $b=0$ but $\{a,b\}\ne0$.
If we set in (\ref{EQ9})  $a=0$, then it takes the following form
\renewcommand{\theequation}{\arabic{equation}${}'$}
\setcounter{equation}{17}
\begin{gather}
\boldsymbol u_t=\boldsymbol u_3 +\frac{3}{2}\left(\frac{\tilde
 u_{[2,2]} }{b}-\frac{(\tilde  u_{[0,2]}+b)^2 }{b\tilde  u_{[0,0]}}+
\frac{\big(\tilde  u_{[0,0]}\tilde  u_{[1,2]}-\tilde  u_{[0,1]}
(\tilde  u_{[0,2]}+b)\big)^2 }{b \xi \tilde  u_{[0,0]}}
- u_{[1,1]} \right)\boldsymbol u_1\nonumber\\
\phantom{\boldsymbol u_t=}{}+ 3u_{[1,2]}\boldsymbol u,\label{EQ9a}
\end{gather}
where $ \xi=\tilde  u_{[0,0]}(b-\tilde  u_{[1,1]})+\tilde  u_{[0,1]}^2$.
If we set in (\ref{EQ10})  $a=0$ and then $b=0$, then the equation is reduced to the following form
\begin{gather}
\boldsymbol u_t=\boldsymbol u_3+3\left(\frac{\tilde u_{[0,1]}}{\tilde u_{[0,0]}}+\frac{\tilde u_{[0,1]}\tilde u_{[0,2]}-\tilde u_{[0,0]}\tilde u_{[1,2]}}{\xi}\right)(\boldsymbol u_2+ u_{[1,1]}\boldsymbol u)+
3 u_{[1,2]}\boldsymbol u\nonumber\\
\phantom{\boldsymbol u_t=}{}+\frac{3}{2}\left(\frac{\tilde u_{[0,0]}\tilde u_{[2,2]}}{\xi}
-\frac{(\xi+\tilde u_{[0,0]}\tilde u_{[0,2]})^2}{\xi \tilde u_{[0,0]}^2}\right)\boldsymbol u_1,\qquad
\xi=\tilde  u_{[0,0]} \tilde  u_{[1,1]}-\tilde  u_{[0,1]}^2. \label{EQ12}
\end{gather}
\end{remark}
\begin{remark} In the classifying process
 we did not consider the isotropic equations because they were found earlier in \cite{M-S1}.
Nevertheless, all equations (\ref{EQ1})--(\ref{EQ11})
admit the reduction $\tilde u_{[i,j]} =k u_{[i,j]}$ that will be referred as  the isotropic reduction.
Each integrable isotropic equation can be obained from the list
 (\ref{EQ1})--(\ref{EQ11}). The diagram of the isotropic reductions takes the
following form:\\[2mm]
\centerline{\framebox{
\unitlength=1mm
\special{em:linewidth 0.4pt}
\linethickness{0.4pt}
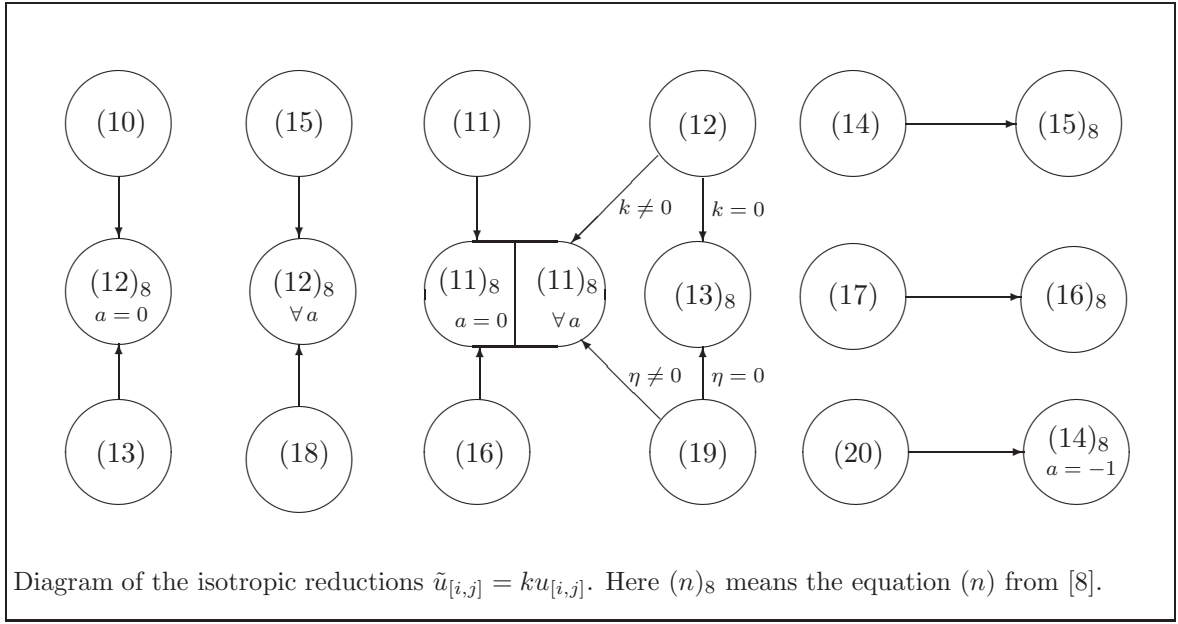
\begin{picture}(149.33,80)(5,60)
\put(15.00,125.00){\circle{14.00}}
\put(15.00,102.67){\circle{14.00}}%
\put(15.00,81.33){\circle{14.00}}
\put(39.50,102.67){\circle{14}}
\put(39.0,80.33){\circle{14.00}}
\put(39.00,125.0){\circle{14.00}}
\put(15.33,125.00){\makebox(0,0)[cc]{(\ref{EQ1})}}
\put(15,118){\vector(0,-1){8.2}}
\put(15.33,103.67){\makebox(0,0)[cc]{$(12)_8$}}
\put(15.37,99.67){\makebox(0,0)[cc]{{\scriptsize$a=0$}}}
\put(15.33,81){\makebox(0,0)[cc]{(\ref{EQ4})}}
\put(39.00,125.00){\makebox(0,0)[cc]{(\ref{EQ6})}}
\put(39.67,81.00){\makebox(0,0)[cc]{(\ref{EQ9})}}
\put(39.67,99.67){\makebox(0,0)[cc]{{\scriptsize$\forall\, a$}}}
\put(39.67,103.67){\makebox(0,0)[cc]{$(12)_8$}}
\put(39,87.4){\vector(0,1){8.2}}
\put(39,118){\vector(0,-1){8.2}}
\put(15,88.37){\vector(0,1){7.4}}
\put(62.67,125.00){\circle{14.00}}
\put(67.67,102.33){\oval(24,14)[]}
\put(67.67,95.33){\line(0,1){14}}
\put(62.67,81.33){\circle{14.00}}
\put(92.00,102.33){\circle{14}}
\put(92.67,81.33){\circle{14.00}}
\put(92.67,125){\circle{14}}
\put(62.67,125){\makebox(0,0)[cc]{(\ref{EQ2})}}%
\put(61.67,104.00){\makebox(0,0)[cc]{$(11)_8$}}
\put(74.67,104.00){\makebox(0,0)[cc]{$(11)_8$}}
\put(59.67,98.00){{\scriptsize$a=0$}}
\put(72.67,98.00){{\scriptsize$\forall\,a$}}
\put(63.00,81){\makebox(0,0)[cc]{(\ref{EQ7})}}
\put(62.67,118.00){\vector(0,-1){8.6}}
\put(63,88.4){\vector(0,1){6.9}}
\put(92.67,124.63){\makebox(0,0)[cc]{(\ref{EQ3})}}
\put(93.00,102.00){\makebox(0,0)[cc]{$(13)_8$}}
\put(92.67,81){\makebox(0,0)[cc]{(\ref{EQ10})}}
\put(86.33,91.33){\makebox(0,0)[cc]{{\scriptsize$\eta\ne0$}}}
\put(97.33,91.33){\makebox(0,0)[cc]{{\scriptsize$\eta=0$}}}
\put(92.67,88.33){\vector(0,1){7.0}}
\put(92.67,117.70){\vector(0,-1){8.4}}
\put(97.33,113.67){\makebox(0,0)[cc]{{\scriptsize$k=0$}}}
\put(87.0,85.7){\vector(-1,1){10.53}}
\put(86.70,120.7){\vector(-1,-1){11.53}}
\put(85.00,113.67){\makebox(0,0)[cc]{{\scriptsize$k\ne0$}}}
\put(112.67,125.00){\circle{14.00}}
\put(141.33,125.00){\circle{14.00}}
\put(142.00,101.67){\circle{14.00}}
\put(112.67,102.00){\circle{14.00}}
\put(113.00,81.33){\circle{14.00}}
\put(142.33,81.33){\circle{14.00}}
\put(119.67,125.00){\vector(1,0){14.67}}
\put(119.67,102.00){\vector(1,0){15.33}}
\put(120.00,81.33){\vector(1,0){15.33}}
\put(113.00,124.67){\makebox(0,0)[cc]{(\ref{EQ5})}}
\put(141.33,124.67){\makebox(0,0)[cc]{$(15)_8$}}
\put(112.67,102.00){\makebox(0,0)[cc]{(\ref{EQ8})}}
\put(142.33,102.00){\makebox(0,0)[cc]{$(16)_8$}}
\put(113.33,81.00){\makebox(0,0)[cc]{(\ref{EQ11})}}
\put(143,79){\makebox(0,0)[cc]{{\scriptsize$a=-1$}}}
\put(142.67,83.00){\makebox(0,0)[cc]{$(14)_8$}}
\put(1,63.00){{\small Diagram of the isotropic reductions  $\tilde u_{[i,j]} =k u_{[i,j]}$.
 Here $(n)_8$ means the equation $(n)$ from \cite{M-S1}.}}
\end{picture}}
}

\medskip

We stress that the equation (\ref{EQ10}) with $\eta\ne0$ is reduced
to (11) from~\cite{M-S1}. But its reduction under  $\eta=0$ --- (\ref{EQ12}), on the contrary,
is reduced to the vector Schwartz--KdV equation (13) from~\cite{M-S1}.
Hence the properties of the equations (\ref{EQ10}) and  (\ref{EQ12})
are essentially different.
\end{remark}
\begin{remark}  While proving the main theorem we found that all equations  (\ref{EQ1})--(\ref{EQ11})
have non-trivial local conserved densities of the orders 2, 3, 4 and 5.
All these densities can be obtained from the formula (\ref{rekkur}).
For example, the equation (\ref{EQ2}) have the following canonical conserved densities:
\begin{gather*}
\rho_0=\frac{u_{[1,2]}}{u_{[1,1]}},\qquad
\rho_1=-\frac{1}{2}\left(\frac{u_{[2,2]}}{u_{[1,1]}}
- \frac{u_{[1,2]}^2}{u_{[1,1]}^2}+\frac{\tilde u_{[1,1]}}{u_{[1,1]}}\right)-D_x\frac{u_{[2,2]}}{u_{[1,1]}}, \\
\rho_2=D_x\left(\frac{u_{[1,3]}}{u_{[1,1]}}+
\frac{3}{2} \frac{u_{[2,2]}}{u_{[1,1]}}-2\frac{u_{[1,2]}^2}{u_{[1,1]}^2}
+\frac{\tilde u_{[1,1]}}{u_{[1,1]}}\right),\quad \dots.
\end{gather*}
\end{remark}

\section{A sketchy proof of the main theorem}

The equation (\ref{3th}) can be rewritten in the form
\[
L\boldsymbol u=0,\qquad L=-D_t+D_x^3+f_2D_x^2+f_1D_x+f_0.
\]
This operator $L$ is used for obtaining the canonical conserved densities by
a technique proposed in \cite{chine}. Motivation and explanation of the technique, for vector equations,
have been presented earlier in~\cite{M-S1}. For more details, see also~\cite{m1}.

Let $\rho$ and $\theta$ be the generating functions for the canonical densities and fluxes correspon\-dingly:
\renewcommand{\theequation}{\arabic{equation}}
\setcounter{equation}{20}
\begin{gather*}
\rho =k^{-1}+\sum_{i=0}^{\infty } \rho_i k^i,
\qquad \theta= k^{-3}+\sum_{i=0}^{\infty } \theta _i k^i,\qquad D_t\rho =D_x\theta.
\end{gather*}
Then, from the equation $L\exp(D_x^{-1}\rho)=0$, the recursion formula follows
\begin{gather}
\rho_{n+2}=\frac{1}{3}\left[\theta_n-f_0 \delta_{n,0} -2 f_2 \rho_{n+1
}-
f_2 D_x\rho_{n} - f_1 \rho_{n}\right] \nonumber\\
\phantom{\rho_{n+2}=}{}-\frac{1}{3}\left[f_2 \sum_{s=0}^{n} \rho_{s} \rho_{n-s}+\sum_{0\leqslant
s+k\leqslant n}\rho_{s} \rho_{k}
\rho_{n-s-k}+3\sum_{s=0}^{n+1}\rho_{s}  \rho_{n-s+1}\right]
\nonumber \\
\phantom{\rho_{n+2}=}{}-D_x\left[\rho_{n+1}+\frac{1}{2}\sum_{s=0}^{n}\rho_{s} \rho_{n-s}+
\frac{1}{3}
D_x \rho_{n}\right],\qquad n\geqslant 0, \label{rekkur}
\end{gather}
where $\delta_{i,j}$ is the Kronecker delta and $\rho_0$,
$\rho_1$ are
\begin{gather} \label{ro01}
\rho_0 = -\frac{1}{3} f_2,\qquad \rho_1 =\frac{1}{9} f_2^2-\frac{1}{3} f_1+\frac{1}{3} D_x f_2.
\end{gather}

The corresponding functions $\theta_i$ can be found from (\ref{cond}).
 The fact that the left-hand sides of~(\ref{cond}) are total
$x$-derivatives imposes rigid restrictions (see below) on the coefficients $f_i$ of (\ref{3th}).

Expressions for the next functions
$\rho_i$ involve, besides  $f_k$, the functions $\theta_j$ with $j\leqslant i-2$. For example
\[
\rho_2 = -\frac{1}{3} f_0+\frac{1}{3} \theta _0-\frac{2}{81} f_2^3+ \frac{1}{9} f_1 f_2-
D_x\left(\frac{1}{9} f_2^2 + \frac{2}{9} D_x f_2 -\frac{1}{3} f_1\right)
\]
and so on.

It is shown in \cite{M-S1} that all
even canonical densities $\rho_{2n}$ are trivial and we have the following strengthened conditions
of the integrability
\begin{gather}
D_t \rho_{2n+1}(u) = D_x \theta_{2n+1}(u),\qquad \rho_{2n}(u) = D_x \sigma_{2n}(u),\qquad n=0,1,\dots  \label{conds}
\end{gather}
instead of (\ref{cond}).

To show how to use the conditions (\ref{conds}), we consider the equations (\ref{EQ}) on $\mathbb S^{n}$.
Obviously,  we have to replace $f_0$ by $f_2 u_{[1,1]} + 3 u_{[1,2]}$ in the formulas (\ref{rekkur}).

\begin{lemma} Suppose the equation \eqref{EQ} on $\mathbb S^{n}$ admits the canonical conserved density $\rho_0$
then the equation has the following form
\begin{gather}
\label{eq1}
\boldsymbol u_t=\boldsymbol u_3-\frac{3}{2} \boldsymbol u_2D_x\ln(g_0)+f_1 \boldsymbol u_1
+\left( 3 u_{[1,2]}-\frac{3}{2}  u_{[1,1]}D_x\ln(g_0)\right)  \boldsymbol u,
\end{gather}
where $g_0$ depends on $\tilde u_{[0,0]}$,  $\tilde u_{[0,1]}$, $\tilde u_{[1,1]}$
and $u_{[1,1]}$, and $f_1$ is  a function of the variables \eqref{dynvar1}.
\end{lemma}

\begin{proof} From (\ref{ro01}) and (\ref{conds})
we have $f_2=D_x\sigma_0$ for some function $\sigma_0$. Since $f_2$ does not depend on
the third order variables, we see that $\sigma_0$ may only depend on
the variables $\tilde u_{[0,0]}$,
$\tilde u_{[0,1]}$, $\tilde u_{[1,1]}$ and $u_{[1,1]}$.
Setting for convenience $\sigma_0=- 3/2\ln(g_0)$ we obtain~(\ref{eq1}).
\end{proof}

\begin{lemma} Suppose the equation \eqref{eq1}
 on $\mathbb S^{n}$ admits the canonical conserved densities~$\rho_1$ and~$\rho_2$,  then
\begin{gather}
f_1=\frac{c_1u_{[2,2]}+c_2\tilde u_{[2,2]}}{g_0}+f_3 u_{[1,2]}^2
+f_4 \tilde u_{[0,2]}^2+f_5 \tilde u_{[1,2]}^2+f_6 u_{[1,2]} \tilde u_{[0,2]} \nonumber\\
\phantom{f_1=}{}+f_7 u_{[1,2]} \tilde u_{[1,2]}+f_8 \tilde u_{[1,2]} \tilde u_{[0,2]}
+f_9 u_{[1,2]}+f_{10} \tilde u_{[0,2]}+f_{11} \tilde u_{[1,2]}+f_{12},\label{f1}
\end{gather}
where $f_i$ are some functions of the variables $\tilde u_{[0,0]}$,  $\tilde u_{[0,1]}$,
$\tilde u_{[1,1]}$ and $u_{[1,1]}$.
\end{lemma}

\begin{proof} To specify the form of the coefficient $f_1$,
 we consider the condition $D_t \rho_1=D_x\theta_1$, where $\rho_1$ is given by (\ref{ro01}).

To simplify the equation $D_t \rho_1=D_x\theta_1$
we use the equivalence relation that will be denoted as~$\sim$. We say that $F_1$ and $F_2$
are equivalent ($F_1\sim F_2$) if $F_1-F_2=D_xF_3$ for some function~$F_3$.
Thus we may write $D_t \rho_1\sim 0$, and this equivalence remains true after adding any $x$-derivative:
$D_t \rho_1+D_x F\sim 0$, $ \forall\, F$.  We call the transformation
$D_t \rho_1\to D_t \rho_1+D_x F$ the equivalence transformation. Using the equivalence 
transformation we can reduce the order of $D_t \rho_1$ step by step. For example, $f(\tilde u_{[0,0]})\tilde u_{[0,1]}
\sim f(\tilde u_{[0,0]})\tilde u_{[0,1]}-\frac 12 D_x\int f(\tilde u_{[0,0]}) d\tilde u_{[0,0]}=0$.

Reducing the order of $D_t \rho_1$ by  the equivalence transformation
we found that  $D_t \rho_1$  is  equi\-va\-lent to a third degree
polynomial of the third  order variables $u_{[i,3]}$, $i=1,2,3$ and  $\tilde u_{[i,3]}$,
$i=0,1,2,3$. As this polynomial must be equivalent to zero,
and it is obvious that any total derivative $D_x F$ is linear with respect to highest order variables, then
the second and third degree terms must vanish. By equating the third degree
terms to zero one can find that all third order derivatives of $f_1$
with respect to second order variables  vanish, and, moreover,
\[
\frac{\partial ^2 f_1g_0}{\partial u_{[2,2]}\partial u_{[i,j]}}=\frac{\partial ^2
f_1g_0}{\partial u_{[2,2]}\partial\tilde u_{[i,j]}}=\frac{\partial ^2 f_1g_0}{\partial\tilde u_{[2,2]}\partial u_{[i,j]}}
=\frac{\partial ^2 f_1g_0}{\partial\tilde u_{[2,2]}\partial\tilde u_{[i,j]}}=0,\qquad \forall \, [i, j].
\]
Integrating all these equations we obtain  (\ref{f1}).
\end{proof}

\begin{remark}  We do not use the equations  $\delta D_t \rho_i(u)/\delta u=0$
because such computations are only possible with the help of
a supercomputer. Unfortunately, our IBM PC with 1 GB RAM does not permit us to do it.
\end{remark}
\begin{lemma} The following three cases are only possible in \eqref{f1}:
\[
{\bf (A)} \ c_1=c_2=0;\qquad {\bf (B)} \ c_1=1,\quad c_2=0; \qquad {\bf (C)} \ c_1=0,\quad c_2=1.
\]
\end{lemma}
\begin{proof} If $c_1\ne 0$, $c_2=0$ we can change $g_0\to c_1 g_0$
and it is equivalent to $c_1=1$. If $c_2\ne 0$, the transformation (\ref{tr})
with  $\lambda =-c_1/c_2$, $\alpha=1/c_2$ gives $c_1=0$ and $c_2=1$.
\end{proof}

Then we specified the functions $g_0, f_3,\dots, f_{12}$
in (\ref{eq1}), (\ref{f1}) in the cases A, B, and C using the
next integrability conditions
(\ref{conds}). These computations are very cumbersome and we can
 not present it in a short article. The result of the computations is the
Theorem~1.

\section{B\"acklund transformations}

To prove integrability of all equations from the list (\ref{EQ1})--(\ref{EQ11})  we present in this section first order
auto-B\"acklund transformations for all equations.
Such transformations involving an arbitrary parameter allow us to build up
both multi-soliton and finite-gap solutions even if the Lax representation
is not known (see \cite{adshyam}). That is why
the existence of an auto-B\"acklund transformation with additional ``spectral'' parameter~$\lambda$ is a convincing
evidence of integrability.

For a scalar evolution equation, a first order auto-B\"acklund transformation is a relation between two
solutions $u$ and $v$ of the same equation and their derivatives $u_x$ and $v_x$. Writing this constraint as
$u_x=\phi(u,v,v_x),$ we can express all derivatives of $u$ in terms of $u,  v,  v_x,
\dots   , v_i, \dots $.
These variables are regarded as independent.

In the vector case, the independent variables are vectors
\begin{gather} \label{vecvar}
\boldsymbol u, \ \boldsymbol v, \ \boldsymbol v_1, \ \boldsymbol v_2, \ \dots, \ \boldsymbol v_i, \ \dots   ,
\end{gather}
and all their scalar products
\begin{gather}
\tilde u_{[0,0]}= \langle \boldsymbol u,   \boldsymbol u\rangle,\qquad
 v_{[i,j]} = (\boldsymbol v_i,   \boldsymbol v_j),\qquad \tilde v_{[i,j]}=
  \langle\boldsymbol v_i,   \boldsymbol v_j\rangle,\nonumber\\
  w_i=(\boldsymbol u,   \boldsymbol v_i),\qquad \tilde w_i=\langle \boldsymbol u,   \boldsymbol v_i\rangle,
  \qquad i,j\geqslant 0.\label{scalvar}
\end{gather}

Following \cite{M-S1}, we consider in this paper special vector auto-B\"acklund transformations of the form
\begin{gather} \label{genbak}
\boldsymbol u_1=h  \boldsymbol v_1+f  \boldsymbol u+g  \boldsymbol v,
\end{gather}
where $f$, $g$ and $h$ are  scalar functions of the variables (\ref{scalvar}) with $i,j\leqslant 1$.
Since $\boldsymbol v$ belongs to the sphere, $(\boldsymbol v,   \boldsymbol v)=1$,
we assume, without loss of generality, that the arguments of $f$, $g$ and
$h$ are
\begin{gather}
\tilde u_{[0,0]},\ \tilde v_{[0,0]},\ w_0,\ \tilde w_0,\ v_{[1,1]},
\  \tilde v_{[0,1]},\ \tilde v_{[1,1]},\ w_1,\ \tilde w_1.
\label{invar}
\end{gather}
Since $(\boldsymbol u,   \boldsymbol u)=1,$ and $(\boldsymbol u,   \boldsymbol u_1)=0,$ it follows from
(\ref{genbak}) that
\begin{gather*}
f=-w_{0}  g-w_{1} h.
\end{gather*}

To find an auto-B\"acklund transformation for
the equation (\ref{nth}), we differentiate (\ref{genbak}) with respect
to $t$ in virtue of (\ref{nth}) and express all vector
and scalar variables in terms of the independent variables
(\ref{vecvar}) and (\ref{scalvar}). By definition of the  B\"acklund
transformation, the expression thus obtained must
be identically equal to zero. Splitting this expression
with respect to the vector variables (\ref{vecvar}) and the  scalar
variables (\ref{scalvar}) different from (\ref{invar}) we derive
an overdetermined system of non-linear PDEs for the functions $f$ and $g$. If the
system has a solution depending on an essential parameter~$\lambda$, this solution gives us the auto-B\"acklund
transformation we are looking for.

We present below the result of our computations.

In the case of  the equation (\ref{EQ1})  the auto-B\"acklund transformation reads as follows
\begin{gather*}\label{BT1}
{\boldsymbol u_x+\boldsymbol v_x}=2 \frac{(\boldsymbol u,\boldsymbol v_x)(\boldsymbol u
+\boldsymbol v)+f  (\boldsymbol v-(\boldsymbol u,\boldsymbol v) \boldsymbol u )}{(\boldsymbol u+\boldsymbol v)^2},
\end{gather*}
 where $f^2=\langle\boldsymbol u+\boldsymbol v, \boldsymbol u+\boldsymbol v\rangle
 +\lambda  (\boldsymbol u+\boldsymbol v)^2$.

Next two equations, (\ref{EQ2}) and (\ref{EQ3}),
are integrable on $\mathbb R^n$ not only on $\mathbb S^n$. In fact, $f_1$ and $f_2$ do not depend
on $u_{[0,i]}$. Hence
\[
D_t u_{[i,j]}=\left(D_x^i(\boldsymbol u_3+f_2\boldsymbol u_2+f_1\boldsymbol u_1),
\boldsymbol u_j\right)+\left(\boldsymbol u_i,D_x^j(\boldsymbol u_3+f_2\boldsymbol u_2+f_1\boldsymbol u_1)\right)
\]
do not depend on $u_{[0,i]}$ for any $i,j>0$.
This implies that all $\rho_n$ and $\theta_n$ for (\ref{EQ2}) and (\ref{EQ3}) do not depend on
$u_{[0,i]}$ (see (\ref{ro01}) and (\ref{rekkur})).
This means that the conditions $D_t\rho_n=D_x\theta_n$
are valid both on $\mathbb R^n$ and on $\mathbb S^n$.

The equations (\ref{EQ2}) and (\ref{EQ3}) on $\mathbb R^n$ have
the auto-B\"acklund transformations of the form
\begin{gather*}
\boldsymbol u_x=F\left(\frac{w_1-v_{[0,1]}}{\varphi} (\boldsymbol u-\boldsymbol v)-\boldsymbol v_x \right),\label{BT2}
\end{gather*}
where $\varphi=\frac 1 2(\boldsymbol u-\boldsymbol v)^2$. The function $F$ reads as
\[
F=\sqrt{\vphantom{\rule{0mm}{2ex}} \frac{\lambda
\varphi- \tilde \varphi}{v_{[1,1]}}} -1,\qquad \tilde \varphi=\tilde u_{[0,0]}+\tilde v_{[0,0]}-2 \tilde w_0
\]
{\samepage for (\ref{EQ2}) and as
\[
F=\frac{\lambda \varphi+\tilde w_0 }{\tilde v_{[0,0]}}
\left(\sqrt{1+\tilde v_{[0,0]} v_{[1,1]}^{-1}} \sqrt{1-\frac{\tilde u_{[0,0]} \tilde v_{[0,0]}}
{(\lambda\varphi+\tilde w_0)^2 }}-1\right)
\]
for (\ref{EQ3}). On $\mathbb S^n$ we have $v_{[0,1]}=0$ and $\varphi=1-w_0$.}

The equations (\ref{EQ4}) and (\ref{EQ5})
have the auto-B\"acklund transformations of the form
\begin{gather*}
\boldsymbol u_x=f\left[\boldsymbol v_x- w_1\boldsymbol u+
\left( \frac{ f \tilde v_{[0,1]}-\tilde w_1 }{f\tilde v_{[0,0]}
-\tilde w_0 } +g \right)(w_0 \boldsymbol u-\boldsymbol v ) \right],\qquad f^2=\frac{\tilde u_{[0,0]}}
{\tilde v_{[0,0]}},\label{BT4-5}
\end{gather*}
where $g=\lambda$ for  (\ref{EQ5}) and $g$ satisfies, for  (\ref{EQ4}), the following equation
\[
g^2 = a \tilde v_{[0,0]}\frac{f^2-2 f w_0+1}{(f \tilde v_{[0,0]}-\tilde w_0)^2}
+\frac{\lambda  f \tilde v_{[0,0]}}{f \tilde v_{[0,0]}-\tilde w_0}.
\]

The auto-B\"acklund transformation for the equation (\ref{EQ6}) is defined by the following equation
\begin{gather*}
\boldsymbol u_x=f\left(\boldsymbol v_x-\boldsymbol u w_1 +\frac{(\lambda w_1+h) (\boldsymbol v-\boldsymbol u w_0)}
{f \tilde v_{[0,0]}-\tilde w_0}\right),\qquad f^2=\frac{\tilde u_{[0,0]}}{\tilde v_{[0,0]}},
\label{BT6}
\end{gather*}
where
\[
h^2=(\tilde v_{[0,0]}+v_{[1,1]})\left(\lambda ^2-(\lambda  w_0-\tilde w_0+f \tilde v_{[0,0]})^2\right).
\]

The auto-B\"acklund transformations for the equations (\ref{EQ7}), (\ref{EQ8}) and (\ref{EQ9}) take the form
\begin{gather*}
\boldsymbol u_x=F (\boldsymbol v_x+g\boldsymbol v- (w_1+g w_0)\boldsymbol u), \label{BT6-9}
\end{gather*}
where
\begin{gather*}
F= h+f ,\qquad f^2=\frac{\tilde u_{[0,0]}}{\tilde v_{[0,0]}},\qquad
 g=-\frac{f\tilde v_{[0,1]}-\tilde w_1 }{f\tilde v_{[0,0]}-\tilde w_0 },\\
h^2=\frac{2}{3} \tilde v_{[0,0]} \frac{a(f^2 +1-2f w_0) -\lambda (\tilde u_{[0,0]}-\tilde w_0 f)  }
{\tilde v_{[0,0]}\tilde v_{[1,1]}-\tilde v_{[0,1]}^2}
\end{gather*}
for  (\ref{EQ7});
\begin{gather*}
F=f-\lambda \frac{\tilde v_{[0,1]}\tilde w_0 - \tilde w_1\tilde v_{[0,0]}}
{\tilde v_{[0,0]}\tilde v_{[1,1]}-\tilde v_{[0,1]}^2 },\qquad
 f^2=\frac{\tilde u_{[0,0]}}{\tilde v_{[0,0]}},\qquad
  g=-\frac{f\tilde v_{[0,1]}-\tilde w_1 }{f\tilde v_{[0,0]}-\tilde w_0 }
\end{gather*}
for  (\ref{EQ8});
\begin{gather*}
g =\frac{1}{\tilde v_{[0,0]}}\left[
\Big(\big( ( 1+h \tilde w_0)^2-\tilde u_{[0,0]}\tilde v_{[0,0]} h^2\big)
\big( \tilde v_{[0,1]}^2+\tilde v_{[0,0]} (a\tilde v_{[0,0]}+b-\tilde v_{[1,1]})\big)\Big)^{1/2}\right.\\
\phantom{g =}{}\left.\vphantom{\Big)^2}-\tilde v_{[0,1]}
-h(\tilde v_{[0,1]}\tilde w_0-\tilde w_1\tilde v_{[0,0]})\right],\\
F=f,\qquad  f^2=\frac{a\tilde u_{[0,0]}+b}{a \tilde v_{[0,0]}+b},\qquad
 h=\frac{\lambda }{(a \tilde v_{[0,0]}+b) f-a\tilde w_0-b w_0}
\end{gather*}
for  (\ref{EQ9}).

The equation (\ref{EQ10}) has the following auto-B\"acklund transformation
\begin{gather}\label{BT10}
\boldsymbol u_x=F\left(
\boldsymbol v_x-w_1\boldsymbol u
+\frac{\tilde w_1+f \tilde v_{[0,1]} }{\tilde w_0+f \tilde v_{[0,0]} }(w_0\boldsymbol u-\boldsymbol v)\right),
\end{gather}
where
\begin{gather*}
F=\phi^{-1}\left(g+\sqrt{g^2-\phi \psi }  \coth{\chi }\right) ,\qquad
g=\lambda(\tilde w_0+f\tilde v_{[0,0]}) -a f\tilde v_{[0,0]}+bw_0,\\
\sinh^2\chi =\frac{\tilde v_{[0,0]} \tilde v_{[1,1]} -\tilde v_{[0,1]}^2}{\tilde v_{[0,0]} \phi },
\qquad f^2=\frac{\tilde u_{[0,0]}}{\tilde v_{[0,0]}},\qquad
\phi=a\tilde v_{[0,0]}+b,\qquad \psi=a\tilde u_{[0,0]}+b.
\end{gather*}
If we set here $a=0$, $\lambda \to b\lambda$ and then $b=0$, we obtain the following expression for $F$:
\[
F=q+\sqrt{q^2-1} ,\qquad q=\lambda(\tilde w_0+f\tilde v_{[0,0]})+w_0.
\]
But it is a trivial solution because (\ref{BT10}) is ``auto-B\"acklund
transformation'' for (\ref{EQ12}) when $F=F(\tilde u_{[0,0]},
\tilde v_{[0,0]},w_0,\tilde w_0)$ is an arbitrary solution of a
quasilinear system of the three first order equations.  (V.~Sokolov and
 A.~Meshkov found for the vector isotropic Schwartz--KdV equation that
  an ``auto-B\"acklund transformation'' containing an arbitrary function
is equivalent to a point transformation. Unfortunately, formula~(49) in~\cite{M-S1}
also gives a false auto-B\"acklund  transformation.)

The true auto-B\"acklund transformation for the anisotropic
Schwartz--KdV equation (\ref{EQ12}) has the following form
\begin{gather*}\label{BT12}
\boldsymbol u_x=\frac{\lambda \tilde v_{[0,0]}}{\zeta}\left(
(\tilde w_0+f \tilde v_{[0,0]})(\boldsymbol v_x-w_1\boldsymbol u)+(\tilde w_1+f \tilde v_{[0,1]} )(w_0\boldsymbol u-\boldsymbol v)\right),
\end{gather*}
where
\[
f^2=\frac{\tilde u_{[0,0]}}{\tilde v_{[0,0]}},\quad \zeta =\tilde v_{[0,0]} \tilde v_{[1,1]} -\tilde v_{[0,1]}^2.
\]
This transformation is reduced in the isotropic limit to the true
auto-B\"acklund transformation for isotropic Schwartz-KdV equation  that was
pesented in~\cite{M-S2}.

Finally, the auto-B\"acklund transformation for the equation (\ref{EQ11}) reads as follows
\begin{gather*}\label{BT13}
\boldsymbol u_x=F\big(\boldsymbol v_x-w_1\boldsymbol u+g(\boldsymbol v - w_0\boldsymbol u)\big),
\end{gather*}
where
\begin{gather*}
F=\frac{1}{\tilde v_{[0,0]}}\left(\lambda \frac{\tilde v_{[0,0]}
\tilde w_1-\tilde v_{[0,1]} \tilde w_0}{\nu +\tilde v_{[0,0]}}-h\right),\qquad
h^2=\tilde u_{[0,0]} \tilde v_{[0,0]}+\lambda^2(\tilde w_0^2-\tilde u_{[0,0]} \tilde v_{[0,0]}),\\
g=(\lambda^2 - 1)\frac{\tilde v_{[0,0]} \tilde w_1-\tilde v_{[0,1]} \tilde w_0}
{\tilde v_{[0,0]}(h+\tilde v_{[0,0]})}- \frac{\tilde v_{[0,1]}+\lambda (\nu+\tilde v_{[0,0]})}{\tilde v_{[0,0]}},\qquad
\nu^2=\tilde v_{[0,1]}^2+\tilde v_{[0,0]}^2 - \tilde v_{[0,0]} \tilde v_{[1,1]}.
\end{gather*}

\section{Concluding remarks}
Each of the  presented B\"acklund transformations
is reduced to the true B\"acklund transformation under the isotropic
reduction $\tilde u_{[i,j]}=k u_{[i,j]}$ in
accordance with the diagram of the the isotropic reductions. It is convincing evidence that  all
B\"acklund transformations are real.

We do not know any applied problems that lead to one of the equations
from our list. If such problem emerge in future, it will
be interesting to find the Lax representation for the corresponding equation.

The presented B\"acklund transformations can be used, first, for obtaining the soliton-like solutions
(see~\cite{Bal1}) and, secondly, for constructing the superposition formulas and new discrete integrable systems.

\subsection*{Acknowledgements}
This research was supported by RFBR grant 05-01-96403. We also grateful to V.V.~Sokolov for many stimulative
discussions.

\LastPageEnding

\end{document}